
\magnification=\magstep1
  
\font\secfont=cmbx10 scaled\magstep1
\newbox\leftpage \newdimen\fullhsize \newdimen\hstitle \newdimen\hsbody
\hoffset=0.0truein \voffset=0.20truein \hsbody=\hsize \hstitle=\hsize
\tolerance=1000\hfuzz=2pt \baselineskip=16pt plus 4pt minus 2pt
\global\newcount\secno \global\secno=0
\def\section#1{\global\advance\secno by1 \xdef\secsym{\ifcase\secno
\or A\or B\or C\or D\or E\or F\or G\or H\or I\or J\fi }
\bigbreak\medskip\noindent\underbar{\secfont\secsym.~#1}
\par\nobreak\medskip\nobreak}
\global\newcount\meqno \global\meqno=1
\def\eqn#1#2{\xdef #1{(\the\meqno)}\global\advance\meqno by1 %
$$#2\eqno#1$$}
\global\newcount\refno \global\refno=1 \newwrite\rfile
\def\ref#1#2{[\the\refno]\nref#1{#2}}%
\def\nref#1#2{\xdef#1{[\the\refno]}%
\ifnum\refno=1\immediate\openout\rfile=refs.tmp\fi%
\immediate\write\rfile{\noexpand\item{[\the\refno]\ }#2}%
\global\advance\refno by1}

\centerline{\secfont Dynamic Scaling Phenomena in Growth Processes}
\bigskip
\centerline{\it Mehran Kardar}
\par
\noindent{ Department of Physics, Massachusetts Institute of Technology,
Cambridge, MA 02139}
\vfill
\centerline{\bf Abstract}
\bigskip
Inhomogeneities in a deposition process may lead to formation of rough
surfaces.
Fluctuations in the height $h({\bf x},t)$, of the surface (at location ${\bf
x}$ and
time $t$) can be probed directly by scanning microscopy, or indirectly by
scattering.
Analytical or numerical treatments of simple growth models suggest that,
quite generally,  the height fluctuations have a self-similar character; their
average correlations exhibiting a dynamic scaling form,
$
\left\langle\left[ h({\bf x},t)- h({\bf x'},t')\right]^2\right\rangle=
|{\bf x-x'}|^{2\alpha} f\left(|{\bf x-x'}|^z/|t-t'|  \right).
$
The {\it roughness} and {\it dynamic} exponents, $\alpha$ and $z$, are expected
to be {\it universal}; depending only on the underlying mechanism that
generates self-similar scaling. Despite its ubiquitous occurrence in theory
and simulations, experimental confirmation of dynamic scaling has been
scarce. In some cases where such scaling has been observed, the exponents are
different from those expected on the basis of analysis or numerics.
I shall briefly review the theoretical foundations of dynamic scaling, and
suggest possible reasons for discrepancies with experimental results.
\vfill
\vfill
\centerline {For the Proceedings of the }
\par\centerline{\it Fourth International Conference on Surface X-Ray and
Neutron Scattering}
\par\centerline{Lake Geneva, June 1995}
\par\centerline{ To be published in a special issue of Physica B}
\vfill\eject
\section{Dynamic Scaling}
The growth of films by deposition is clearly of great technological interest.
It has recently been recognized that such growth processes also
pose important issues in the physics of non-linear and complex systems.
Rather crudely, we can distinguish between three types of growth morphologies:

\noindent{\bf 1.} {\it Layer by layer} growth is naturally most desired from
the technological stand-point.  From a theoretical perspective, this type
of growth is not stable in the presence of inhomogeneities in the deposition
beam. However, this is only a statement about asymptotic behavior, as
it is clearly possible to grow many layers in this mode.

\noindent{\bf 2.} {\it Unstable} growth occurs when the selected orientation
of the substrate cannot be maintained. The instability is usually manifested by
the  formation  of mounds, or other macroscopic features on the surface.
Another type of instability results from growth that is controlled by a
diffusive
field. This instability may lead to formation of fractal aggregates. This is
probably unrelated to growth by deposition and will not be discussed here.

\noindent{\bf 3.} {\it Self--affine} surfaces appear in a growth mode
that is intermediate between the above. The average orientation of the
surface is maintained, but it becomes rough; the amount of roughness
grows with time and/or scale of observation in a self--similar fashion.

The most economical way to characterize self--affine roughness is
by appealing to a dynamic scaling form\ref
\rFV{F. Family and T. Vicsek, J. Phys.  A {\bf 18}, L75 (1985).}:
If the height of the surface at location ${\bf x}$, at time $t$, is described
by
a function $h({\bf x},t)$, its average correlations may satisfy the scaling
form
\eqn\eDS{g\left( {\bf r},t \right)\equiv\left\langle \left[
h\left( {\bf x},\tau \right)-h\left( {\bf x}+{\bf r},\tau+t \right)\right]^2
\right\rangle
=|{\bf r}|^{2\alpha}f\left( {t \over |{\bf r}|^{z}} \right).}
Similar scaling forms are frequently encountered in the theory of critical
phenomena. Family
and Vicsek\rFV\ first proposed such scaling in the context of interface growth
based on numerical results. There is now a firm theoretical basis for such
behavior. The {\it dynamic exponent} $z$ describes the evolution of correlated
regions with time: initially different parts of the surface are independent;
but
regions of correlated roughness form over time, their size growing as
$\xi(t)\propto t^{1/z}$. In each correlated region, the width of surface grows
as the observation scale raised to the {\it roughness exponent} $\alpha$.
Thus, the overall width of the surface initially grows as $t^\beta$ with
$\beta=\alpha/z$, until it saturates at $L^\alpha$ were $L$ is the sample size.
\section{Discrete Models}
A large number of numerical models of growth, of varying levels of
sophistication,
have been developed\ref
\rMeakin{For a review see, P. Meakin, Phys. Rep. {\bf 235}, 191 (1993).}.
Here, I shall briefly describe some of the simplest versions.

\noindent{\it Random Deposition:} Particles are dropped randomly over
deposition sites, and stick to the top of the pre-existing column on that
site\ref
\rWGJ{J.D. Weeks, G.H. Gilmer, and K.A. Jackson, J. Chem. Phys.
{\bf 65}, 712 (1976).}.
The height of each column thus performs an independent random walk,
such that
\eqn\eRDC{\left\langle h_i(t) \right\rangle=vt,\qquad\qquad
{\rm and}\qquad\qquad
\left\langle\left(  h_i(t)-vt \right)^2 \right\rangle=Dt,}
where $v$ is the average deposition rate per column, and $D$ denotes its
fluctuations. Since the columns are independent, the correlation length
does not grow with time, $\xi=1$ and $1/z=0$. This model leads to
unrealistically
rough surfaces whose overall width grow with the exponent $\beta=1/2$
without saturation.

\noindent{\it Sedimentation:} After the dropped particle reaches the top of
a column, it rolls down-hill to neighboring columns until it reaches a local
height minimum, where it stops. The smoothening effects of the rolling
particles
generate correlations in height, and lead to more realistic surfaces.
Since the motion of rolling particles is more or less diffusive, it is likely
that
the correlation length grows as $\xi(t)\propto t^{1/2}$. This value of $z=2$
is supported by numerical simulations\ref
\rFSed{F. Family, J. Phys. A {\bf19}, L441 (1986).},
which find in $d=2$ (i.e. for a one--dimensional substrate), the exponents
$\alpha=0.24\pm 0.01$ and $\beta=0.48\pm0.02$.

\noindent{\it Ballistic Deposition:} Originally introduced for particles moving
in a continuous space\ref
\rVold{M.J. Vold, J. Coll. Sci. {\bf 14}, 168 (1959).},
several variants with discretized particle locations have since been
developed\rFV\rMeakin. The deposited particles are now attached at the
{\it first point of contact} with the growing aggregate. Since the first
contact
may be to the side of a high column, this model leaves behind a finite density
of voids as it grows.

\noindent{\it Restricted Solid on Solid models:} There are no voids, and the
height is a single valued function of position. These models impose a
restriction
on the maximum height difference between neighboring columns\ref
\rKK{J.M. Kim and J.M. Kosterlitz, Phys. Rev. Lett. {\bf 62}, 2289 (1989).}.
A growth attempt that violates such a restriction is rejected. (This can be
regarded as a crude form of desorption.) The last two models are expected to
be in the same universality class, but better numerical values for the
exponents
are found from the latter. The simulations in $d=2$ give $\alpha\approx 1/2$
and $\beta\approx 1/3$, while for $d=3$ they result in $\alpha\approx 0.38$
and $\beta\approx 0.24$.

\noindent{\it ``Realistic models"} attempt to more closely mimic the physical
processes involved in the growth of surfaces\ref
\rWolf{See e.g., D.E. Wolf, in {\it Scale Invariance, Interfaces, and
Nonequilibrium
Dynamics}, edited by A. McKane, M. Droz, J. Vannimenus, and D. Wolf
(Plenum Press, New York, 1995), page 215.}.
These include the detailed motion of the deposited atom, its subsequent
movement
on the surface; incorporation on steps or islands, island nucleation, and
desorption. The main difficulty with simulating all these processes is that
they take place over a wide range of time scales, making the analysis of large
scale roughness quite difficult.
\section{Continuum Equations}
Rather than describing the detailed microscopic evolution of the surface,
continuum equations focus on the (hopefully universal) macroscopic
aspects of its roughness, e.g. the exponents $\alpha$ and $\beta$.
The general philosophy is to examine the evolution of coarse-grained
(hydrodynamic) variables;  in the example at hand, the height function
$h({\bf x},t)$. Whereas in near equilibrium situations the evolution equation
is obtained from variations of an energy functional, such an approach is
not appropriate far away from equilibrium. Here we follow an approach
described in references \ref
\rMK{M. Kardar, in {\it Disorder and Fracture}, edited by J.C. Charmet, S.
Roux, and E. Guyon (Plenum, New York, 1990), p. 3.}\ and \ref
\rITU{M. Kardar, Tr. J. of Phys. {\bf 18}, 221 (1994).}:
The equation of motion is decomposed as
\eqn\eGen{{\partial {h}\over \partial t}=\eta\left( {\bf x},t \right)
+\Phi\left[ h,\nabla h,\nabla^2h,\cdots \right].}

\item{}{\it Particle deposition} is described by the first term. Thus
$\eta\left( {\bf x},t \right)$ is a random function whose mean value gives the
average
particle flux at ${\bf x}$, and its fluctuations represent the shot noise in
deposition.
For simplicity, we shall assume that the noise is uncorrelated at different
sites
and different times, i.e. it is a Gaussian process with
\eqn\eWnoise{\eqalign{
\left\langle \eta\left( {\bf x},t \right) \right\rangle &= v\cr
\left\langle \delta \eta\left( {\bf x},t \right)\delta \eta\left( {\bf x'},t'
\right)\right\rangle
&=D\delta\left( {\bf x-x'} \right)\delta\left( t-t' \right)
}.}

\item{}{\it Surface relaxation} subsequent to deposition is described by the
functional $\Phi$. As we shall see in the following examples, it can depend on
various properties of the height, such as its slope $\nabla h$, or curvature
$\nabla^2 h$. We shall assume that the relaxation is {\it local}, i.e., it can
be adequately described by the first few terms of an expansion of $\Phi$ in $h$
and
its gradients. Which terms can be included in such an expansion are then
determined by the underlying symmetries and conservation laws appropriate
to the dynamics. The basic idea is that any term that is not excluded for
fundamental reasons of symmetry or conservation, will be generically present.

We shall now describe continuum equations for the discrete models
introduced in the previous section.

\noindent{\it Random deposition} with no subsequent relaxation corresponds
to $\Phi=0$. Integrating $\partial_t h=\eta({\bf x},t)$ yields
\eqn\eCRD{h\left( {\bf x},t \right)=\int_0^t dt'\eta({\bf x},t') ,}
from which we can immediately obtain
\eqn\eARD{\left\langle h\left( {\bf x},t \right) \right\rangle=vt,
\qquad{\rm and}\qquad
\left\langle \delta h\left( {\bf x},t \right)\delta h\left( {\bf x'},t \right)
\right\rangle
=Dt\delta\left( {\bf x-x'} \right),}
corresponding to $\beta=1/2$, and a zero correlation length.

\noindent{\it Sedimentation} was originally analyzed by Edwards and
Wilkinson (EW) in ref.\ref
\rEW{S.F. Edwards and D.R. Wilkinson, Proc. R. Soc. Lond.
{\bf A381}, 17 (1982).}.
They concluded that the main relaxational process is proportional to local
curvature, leading to
\eqn\eEW{\partial_t h=\eta({\bf x},t)+\nu\nabla^2 h.}
This linear (diffusion) equation is readily solved in Fourier space, leading to
\eqn\eEWsol{\left\langle |h({\bf k},\omega)|^2 \right\rangle={D\over
\omega^2+\nu^2 k^4}.}
Recasting the above result in real space then leads to the exponents
\eqn\eEWexp{z=2,\qquad\qquad\alpha={3-d\over 2},\qquad\qquad\beta={3-d\over
4},}
in general space dimension $d$. In particular, in $d=2$, $\alpha=1/2$ and
$\beta=1/4$,
while for $d=3$, the mean-square width grows logarithmically in both space and
time.

\noindent{\it Ballistic deposition} is generically a non-linear process. There
is no
a priori reason why the relaxation function $\Phi$ should not depend on the
slope
$\nabla h$.  (A direct dependence on $h$ itself is ruled out by the
translational
symmetry $h\to h+{\rm constant}$, of the underlying dynamics.) By symmetry, the
relaxation should be the same for slopes $\pm\nabla h$, and hence the first
term
in an expansion in powers of slope starts with $(\nabla h)^2$, leading to
\eqn\eKPZ{\partial_t h=\eta({\bf x},t)+\nu\nabla^2 h+{\lambda\over2}
\left( \nabla h \right)^2+\cdots.}
Higher order terms can also be added, but are irrelevant in that they don't
change
the scaling properties. There are several excellent reviews of eq.\eKPZ, known
as the KPZ equation\ref
\rKPZ{M. Kardar, G. Parisi, and Y.-C. Zhang, Phys. Rev. Lett. {\bf 56}, 889
(1986).}.
Since the origin and properties of the nonlinear term are discussed in detail
in these
reviews\ref
\rHHZ{T. Halpin--Healy and Y.-C. Zhang, Phys. Rep. {\bf 254}, 215 (1995).}\ref
\rBS{A.-L. Barabasi and H. E. Stanley, {\it Fractal concepts in surface
growth},
(CUP, Cambridge, 1995).},
I shall not elaborate on them any further. In $d=2$, the non-linear equation
can
in fact be solved exactly, and leads to $\alpha=1/2$ and $\beta=1/3$ (as
compared
to $\beta=1/4$ for the EW equation), in excellent agreement with the numerical
simulations\rFV\rMeakin. There are no exact solutions in $d=3$, but based
on simulations we can estimate $\alpha\approx0.38$ and $\beta\approx 0.24$.
A characteristic signature of eq.\eKPZ\ is the exponent identity
\eqn\eGI{\alpha+{\alpha\over\beta}=2,}
obeyed in almost all simulations. The KPZ equation appears to describe the
asymptotic behavior of most local, random growth processes.
\section{Conservative ``MBE" Models}
We may ask why the KPZ nonlinearity is not present in the model of
sedimentation. The reason is that it is forbidden by a conservation law.
If the growth process does not allow the formation of {\it overhangs}
or {\it voids}, and there is also no {\it desorption}, then all the incoming
flux is incorporated into the growing aggregate. This means that the net mass
of the aggregate, proportional to $H(t)=\int d^dx h({\bf x},t)$, can
only change due to the random deposition; the relaxation processes
should conserve $H(t)$. This immediately implies that the relaxation function
must be related to a surface current, i.e.
\eqn\ePj{\Phi=-\nabla\cdot {\bf j}\left[ \nabla h,\nabla^2 h,\cdots \right].}
The KPZ non-linearity is thus ruled out, as it cannot be written
as the divergence of another function. Following some observations of
Villain\ref
\rVillain{J. Villain, J. Phys. I {\bf 1}, 19 (1991).},
many studies have focused on such conservative models in the context
of MBE growth. The basic idea is that aggregates formed by the MBE process
are typically free from holes and defects, and that the desorption of the
adsorbed particles is negligible. It is thus argued that, at least over some
sizeable pre-asymptotic regime, the relaxation processes should be
conservative.
Some examples of such conservative models are discussed in the remainder
of this section.

\noindent{\it Surface diffusion} currents in equilibrium can be related to
variations
of a chemical potential via ${\bf j}\propto -\nabla \mu$. Since in the
equilibrium
between two phases, the chemical potential is proportional to local curvature
($\mu\propto-\nabla^2 h$), this leads to an equation of motion
\eqn\eSD{\partial_t h=\eta({\bf x},t)-\kappa\nabla^4 h.}
This equation is again linear, and as for eq.\eEWsol\ can be solved by
Fourier transformation to yield,
\eqn\eSDsol{\left\langle |h({\bf k},\omega)|^2 \right\rangle=
{D\over \omega^2+\kappa^2 k^8}.}
The corresponding exponents in real space are,
\eqn\eSDexp{z=4,\qquad\qquad\alpha={5-d\over 2},\qquad\qquad\beta={5-d\over
8}.}
This leads to $\alpha=3/2$ and $\beta=3/8$ in $d=2$, and
$\alpha=1$ and $\beta=1/4$ in $d=3$. Note that a surface remains self--affine,
maintaining a well defined orientation, only as long as $\alpha<1$. The above
large
values of the exponent $\alpha$ indicate a break-down of
validity of the above equation.

\noindent{\it Non-linear ``MBE"} models have been proposed partly to remedy
the break-down of the linear equation for $\alpha\geq1$. One such model starts
with a non-linear chemical potential introduced by Sun, Guo, and Grant\ref
\rSGG{T. Sun, H. Guo, and M. Grant, Phys. Rev. A {\bf 40}, 6763 (1989).},
resulting in
\eqn\eMBE{\partial_t h=\eta({\bf x},t)-\kappa\nabla^4 h+{\lambda' \over 2}
\nabla^2\left( \nabla h \right)^2.}
Despite its non-linear form, eq.\eMBE\ can in fact be analyzed to yield the
exact exponents\ref
\rLD{Z.W. Lai and S. Das Sarma, Phys. Rev. Lett. {\bf 66}, 2348 (1991).}
\eqn\eMBE{z={7+d\over3},\qquad\qquad\alpha={5-d\over 3},\qquad
\qquad\beta={5-d\over 7+d}.}
In particular $\alpha=2/3$ and $\beta=1/5$ in $d=3$. However, in
non-equilibrium
circumstances, there is no reason for the surface current to be derivable from
a chemical potential. Removing this restriction allows the inclusion of other
non-linearities\ref
\rSP{M. Siegert and M. Plischke, Phys. Rev. E {\bf 50}, 917 (1994).},
such as $\nabla\left( \nabla h \right)^3$, which are in fact more relevant.
More importantly, non-equilibrium currents should generically include a
term proportional to $\nabla h$, which is the dominant gradient as discussed
next.

\noindent{\it Diffusion bias} refers to such generic non-equilibrium currents
that are proportional to the local surface slope\ref
\rKPZ{J. Krug, M. Plischke, and M. Siegert, Phys. Rev. lett. {\bf 70}, 3271
(1993).}.
One possible origin of such currents is in the Schwoebel barriers\ref
\rSB{R.L. Schwoebel, J. Appl. Phys. {\bf 40}, 614 (1969).}:
Atoms on a stepped surface are easily incorporated at the step to a higher
ledge,
but are reflected by a barrier towards jumping to a lower ledge\ref
\rBCF{W.K. Burton, N. Cabrera, and F.C. Frank, Phyl. Trans. R. Soc. Lond.
{\bf A243}, 299 (1951).}.
This sets up a net up-hill current\rVillain\ ${\bf j}=\nu'\nabla h$, leading to
an
equation of motion
\eqn\eBCF{\partial_t h=\eta({\bf x},t)-\nu'\nabla^2 h+\cdots.}
This equation leads to an unstable growth of fluctuations, and therefore higher
order terms are necessary to ensure stability. For example, Johnson et al\ref
\rSLUG{M.D. Johnson, C. Orme, A.W. Hunt, D. Graff, J. Sudijono, L. M. Sander,
and
B.G. Orr, Phys. Rev. Lett. {\bf 72}, 116 (1994).}\
have proposed the following non-linear equation
\eqn\eSLUG{{\partial {h}\over \partial t}=\eta({\bf x},t)-\nabla\left(
{\nu'\nabla h \over
1+\left( \nabla h \right)^2} \right)-\kappa\nabla^4h.}
The instabilities in this equation develop into a complex array of mounds
dubbed
as SLUGs (Super Large Unstable Growths). Finally, it is also possible for the
non-equilibrium currents to be oriented down-hill (as in sedimentation), in
which case the behavior is the same as the EW equation \eEW\ discussed earlier.
\section{Discussion}
An excellent review of the experimental research in this subject is provided by
Krim and Palasantzas\ref
\rKP{J. Krim and G. Palasantzas, Int. J. Mod. Phys. B, in press (1995).}.
The observational methods include diffraction (specular or diffuse X--rays,
RHEED,
LEED, HR-LEED, and helium scattering), direct imaging (STM, AFM, SEM, TEM),
and surface adsorption. A variety of metallic (silver, gold, copper, iron),
and other (Si, InP, NbN, polymer) surfaces grown under a host of different
conditions have been examined by such probes.  Some of these surfaces
exhibit unstable growth, while others appear to satisfy self-similarity
scaling.
However, there is usually no clear--cut identification of the exponents with
the theoretical models. Some experiments on gold and silver give roughness
exponents consistent with the KPZ value, but larger values of $\beta$.
Other surfaces give larger values of $\alpha$, consistent with those of the
non-linear
``MBE" equation \eMBE.  The reader is referred to this review article \rKP\ for
the details. Perhaps the following statements at the end of the review are most
revealing of the experimental situation: ``Over \%50 of the experimental work
reported on here was published in the interval from January 1993 to August
1994. The pace of experimental work is clearly accelerating, and rapid
advances in the field can be expected."

Given the discrepancies between experiment and theory, we can also ask if
important elements have been left out of the analysis. The formalism presented
so far deals solely with a single coarse-grained variable, the height
$h({\bf x},t)$. Other variables may play an important role in the evolution of
$h$. For example, in many cases the roughness is intimately related to
formation of micro-crystalline grains. Variations in crystallinity have so far
been left out of the theoretical picture. In principle, one could introduce
an additional ``order parameter" $M({\bf x},t)$ describing the local degree
of crystallinity. Surface relaxation may then depend on this order parameter,
leading to $\Phi\left( \nabla h,\nabla^2 h, M,\nabla M,\cdots \right)$.
We should then also include an additional dynamical equation for the
evolution of $M$. Similar coupled stochastic equations have been proposed
in the context of surfactant mediated growth\ref
\rBK{A.-L. Barabasi and E. Kaxiras, preprint (1995).}.
A simple variation of ballistic deposition which exhibits complex grain
formation as a result of a sublattice symmetry breaking has also been
proposed\ref
\rBC{C.T. Capraro and Y. Bar-Yam,  J. Comp. Mat. Sci. {\bf 1},169 (1993).}\
in the context of growth of diamond films.

Equation \eGen\ can be regarded as representing a complex filter, converting
the white input noise $\eta({\bf x},t)$ to the correlated random function
$h({\bf x},t)$ through the action of $\Phi$. So far, the focus has been on the
relaxation
function $\Phi$, assuming that the input noise is uncorrelated. Not
surprisingly,
correlations in the input noise lead to more correlated surface roughness with
larger values of the exponents $\alpha$ and $\beta$\ref
\rMHKZ{E. Medina, T. Hwa, M. Kardar, and Y.-C. Zhang, Phys. Rev. A {\bf 39},
3053 (1989).}.
Maybe, in view of the rather large exponents observed experimentally, this
point should be further investigated. Starting with $D({\bf k},\omega)\equiv
\left\langle |\eta({\bf k},\omega)|^2 \right\rangle\propto |{\bf k}|^{-2\rho}$,
roughness exponents may be tuned continuously by changing the parameter
$\rho$. This is not satisfactory, as it relies on a rather arbitrarily chosen
exponent for noise correlations.  Here I shall propose another possibility
which is less arbitrary, and may be relevant to some experiments. My choice
of noise correlations is motivated by the scaling phenomena observed in
turbulence, another interesting problem in non-linear physics.

In the case of turbulence, the fluid is stirred at long length scales, setting
up
a {\it Kolmogorov energy cascade}\ref
\rKol{A.N. Kolmogorov, C. R. Acad. Sci. USSR {\bf 30}, 301 (1941);
{\it ibid.} {\bf 32}, 16 (1941).}\
by which energy is transferred to shorter wave-length modes, ultimately
dissipating at a microscopic scale. In the intermediate (inertial) regime, the
energy density follows simple power laws.
We may similarly assume that the deposition noise is correlated over long
distances $\ell$, such that $D({\bf k})$ is large only for $|{\bf k}|<1/\ell$.
Distances intermediate between the atomic scale and $\ell$ are then
analogous to the inertial regime in turbulence. Adapting the arguments of
Kolmogorov to the KPZ equation leads to
\eqn\eTur{z\approx {4\over3},\qquad\qquad\alpha\approx {2\over 3},\qquad
\qquad\beta\approx {1\over 2},}
in all dimensions.
(Because there is no conservation of a corresponding energy for the KPZ
equation in $d\neq1$, the arguments leading to eq.\eTur\ are considerably
less compelling than those in turbulence. Hence, the above exponents
should be regarded as a first approximation.)
The exponent $\beta=1/2$ is similar to that of random deposition.
Indeed, an early experiment on recrystallization of amorphous GaAs films\ref
\rLNdA{C. Licoppe, Y.I. Nissim, C. d'Anterroches, Phys. Rev. B {\bf 37}, 1287
(1988).},
obtained an exponent of $\beta\approx 0.50$ for highly correlated surfaces.
I will thus conclude this brief review with a question for the experiments:
Is it possible to measure directly the correlations in the deposition noise?

{\bf Acknowledgements:} This research is supported by the NSF through grant
number DMR-93-03667.
\vfill\eject\immediate\closeout\rfile
\centerline{{\bf References}}\bigskip
{\catcode`\@=11\escapechar=`  \input refs.tmp\vfill\eject}
\end